\documentclass[%
 aip,
 amsmath,amssymb,
 reprint,%
]{revtex4-1}

\usepackage{graphicx}
\usepackage{dcolumn}
\usepackage{bm}

\usepackage{lineno,hyperref}
\usepackage{bookmark}
\usepackage{stmaryrd}
\usepackage{amssymb}
\usepackage{amsmath}

\usepackage[utf8]{inputenc}
\usepackage[T1]{fontenc}
\usepackage{mathptmx}

\begin{document}

\preprint{AIP/123-QED}

\title{Effect of O-doping or N-vacancy on the structural, electronic and magnetic properties of MoSi$_2$N$_4$ monolayer}


\author{Yan-Tong Bian}
\affiliation{School of Physical Science and Technology, Tiangong University, Tianjin 300387, China}
\author{Guang-Hua Liu}
\email{liuguanghua@tiangong.edu.cn}
\author{Sheng-Hui Qian}
\author{Xin-Xin Ding}
\author{Jia-Xi Gao}
\author{Hao-Xuan Liu}

\date{\today}

\begin{abstract}
In this letter, the effect of four types of defects (O$_N$$_{out}$, O$_N$$_{in}$, V$_N$$_{out}$ and V$_N$$_{in}$) on the structural, electronic and magnetic properties of MoSi$_2$N$_4$ monolayer were investigated using first-principles calculations.
The calculated results reveal that all the four types of defects lead to structural distortions around the O-dopant or N-vacancy, and thereby change the lattice parameter $a'$ and monolayer height $h$.
Specifically, O$_N$$_{out}$ or O$_N$$_{in}$ increases the lattice parameter, but V$_N$$_{out}$ or V$_N$$_{in}$ is on the contrary.
O$_N$$_{out}$ or V$_N$$_{out}$ increases the monolayer height, whereas the height decreases for O$_N$$_{in}$ or V$_N$$_{in}$.
Each of the four types of defects has a fundamental effect on the electronic properties of MoSi$_2$N$_4$ monolayer, which can induce a transition from semiconductor to metal.
O$_N$$_{in}$ or V$_N$$_{in}$ plays a vital role in the occurrence of a transition from non-magnetism to ferrimagnetism in MoSi$_2$N$_4$ monolayer.
The effect of biaxial strain on the magnetic properties of the two systems with O$_N$$_{in}$ and V$_N$$_{in}$ was subsequently investigated.
It is found that the total magnetic moments are less sensitive to biaxial strain whereas the local magnetic moments residing on the Mo atoms are increased for the two systems with O$_N$$_{in}$ and V$_N$$_{in}$, as strain increases from -3\% to 10\% and from -9\% to 10\%, respectively.
Furthermore, the magnetic phase transitions between ferrimagnetic and paramagnetic states were found to occur around -4\% strain and within the range of -10\% $\thicksim$ -9\% for the two systems with O$_N$$_{in}$ and V$_N$$_{in}$, respectively.
This study may provide a guidance for the application of MoSi$_2$N$_4$ monolayer in the spintronic and magnetic materials.
\end{abstract}

\maketitle



Two-dimensional (2D) materials, exhibiting physical and chemical properties that are richer than and completely different from the 3D bulk crystals due to the inherent quantum confinement effect along the direction of atomic-layered height, and thus have drawn considerable attentions from the scientific and industrial circles on a global scale in the past few years.
The most typical and also earliest 2D material confirmed experimentally is graphene \cite{Novoselov666}, which possesses unique electrical properties.
Researchers subsequently got inspirations from it and developed a series of new 2D materials with excellent mechanical, thermodynamic, electrical, optical and magnetic properties, opening up another valuable path for the development and investigation of high-performance electronic devices, optoelectronic devices, spintronic devices, as well as energy conversion and storage fields \cite{doi:10.1021/acsnano.6b04898,Baugher2014,Mak2012,doi:10.1021/nl401544y}.

The MoSi$_2$N$_4$ compound, a novel synthetic 2D layered van der Waals (vdW) material without known 3D layered parents fabricated very recently by Hong \textit{et al}. \cite{Hong670} through chemical vapor deposition (CVD) technology, are attracting great interest in the optoelectronics, valleytronics, spintronics and so on.
The synthesis of such material provides a huge opportunity for engineering materials with new attributes and functionality \cite{10.1093/nsr/nwaa190}.
Hong \textit{et al}. studied the MoSi$_2$N$_4$ monolayer film and reported that it is built up by septuple-atomic-layer of N-Si-N-Mo-N-Si-N, and exhibits non-magnetic semiconducting nature, high strength, as well as excellent ambient stability.
They investigated its optical properties and evaluated the band gap via Tauc plot analysis further, and employed density functional theory (DFT) \cite{PhysRev.136.B864,PhysRev.140.A1133} to explain observed phenomena.
It is found that the MoSi$_2$N$_4$ monolayer possesses an indirect band gap of 1.94 eV and displays a high optical transmittance with an average of 97.5 $\pm$ 0.2\% in the visible range, and the spin-orbit coupling (SOC) existing in the valence band (VB) can be used to elucidate the two sub-peaks in the optical absorption spectrum.
Wang \textit{et al}. \cite{wang2020structuredriven} analyzed the crystal structure possessing septuple-atomic-layered characteristic for the MoSi$_2$N$_4$ monolayer and reported that it can be viewed as the insertion of the triple-atomic-layered 2$H$-MoS$_2$-type MoN$_2$ monolayer into the quadruple-atomic-layered $\alpha$-InSe-type Si$_2$N$_2$ monolayer.
The piezoelectric and electronic properties of MoSi$_2$N$_4$ monolayer under strain within the range from -4\% to 4\% were studied by Guo \textit{et al}. \cite{guo2020intrinsic} and found that the biaxial tensile strain can enhance the piezoelectric strain coefficient $d$$_{11}$ and spin-orbit splitting at K point, whereas the band gap decreases with increasing strain.
Guo \textit{et al}. \cite{guo2020tuning} investigated the effect of large biaxial strain ranging from -10\% to 10\% on the electronic structure, transport and piezoelectric coefficients of MoSi$_2$N$_4$ monolayer and reported that as strain increases from -10\% to 10\%, the band gap first increases and then decreases.
Bafekry \textit{et al}. \cite{bafekry2020mosi2n4} explored the structural, mechanical, thermal, electronic, optical and photocatalytic properties of MoSi$_2$N$_4$ monolayer and found that it possesses a good thermoelectric performance, and its work function is very similar to that of phosphorene and MoS$_2$ monolayers.
Kang \textit{et al}. \cite{kang2020second} studied the second harmonic generation (SHG) effects of typical MoSi$_2$N$_4$ phases and reported that the two most typical phases may exhibit different SHG intensities due to the difference in P between the outer and inner layers of the sandwich structure.
Li \textit{et al}. \cite{li2020valleydependent} and Yang \textit{et al}. \cite{yang2020valley} explored the valley-dependent properties of MoSi$_2$N$_4$ monolayer and found that it is a semiconductor with a pair of Dirac-type valleys located at the corners of the hexagonal Brillouin zone and can realize the valley-contrast properties, suggesting that its potential applications in valleytronic and spintronic devices.
Moreover, Cao \textit{et al}. \cite{cao2020twodimensional} investigated the vdW heterostructures composed of MoSi$_2$N$_4$ vertically contacted by graphene and NbS$_2$ monolayers and showed that MoSi$_2$N$_4$/NbS$_2$ contact exhibits an ultralow Schottky barrier height (SBH), and the SBH can be modulated via interlayer distance or external electric fields for MoSi$_2$N$_4$/graphene contact, opening up an opportunity for reconfigurable and tunable nanoelectronic devices.

As is well known, the physical and chemical properties of 2D materials under diverse conditions and requirements can commonly be effectively modulated by using of defect, strain, external electric field \cite{PhysRevB.87.155304,PhysRevMaterials.4.114002,https://doi.org/10.1002/pssb.202000343,SHUKLA2020137717,LIN2020146960,WANG2021148098,doi:10.1021/jz301339e,HU201569,doi:10.1063/1.5097264,BAFEKRY2020371} and so on, which provides convenience for investigating the optical, electrical, magnetic and topological properties etcetera closely related to the electronic structure.
Despite an increasing theoretical exploration toward MoSi$_2$N$_4$ monolayer recently, to date the effect of defect on its physical properties has not yet been investigated.
Besides, such work is significant to understand the natures of MoSi$_2$N$_4$ compound and develop its practical applications further.

In view of the facts, herein we targeted to view the structural, electronic and magnetic properties of MoSi$_2$N$_4$ monolayer induced by the defects with O-doping and N-vacancy using first-principles approach based on DFT.
The electronic and magnetic properties were studied through GGA + SOC and GGA + Spin Polarization calculations, respectively.
The calculated results of intrinsic MoSi$_2$N$_4$ monolayer were also given in this letter for comparison.

\begin{figure*}
	\includegraphics[height=12.0cm]{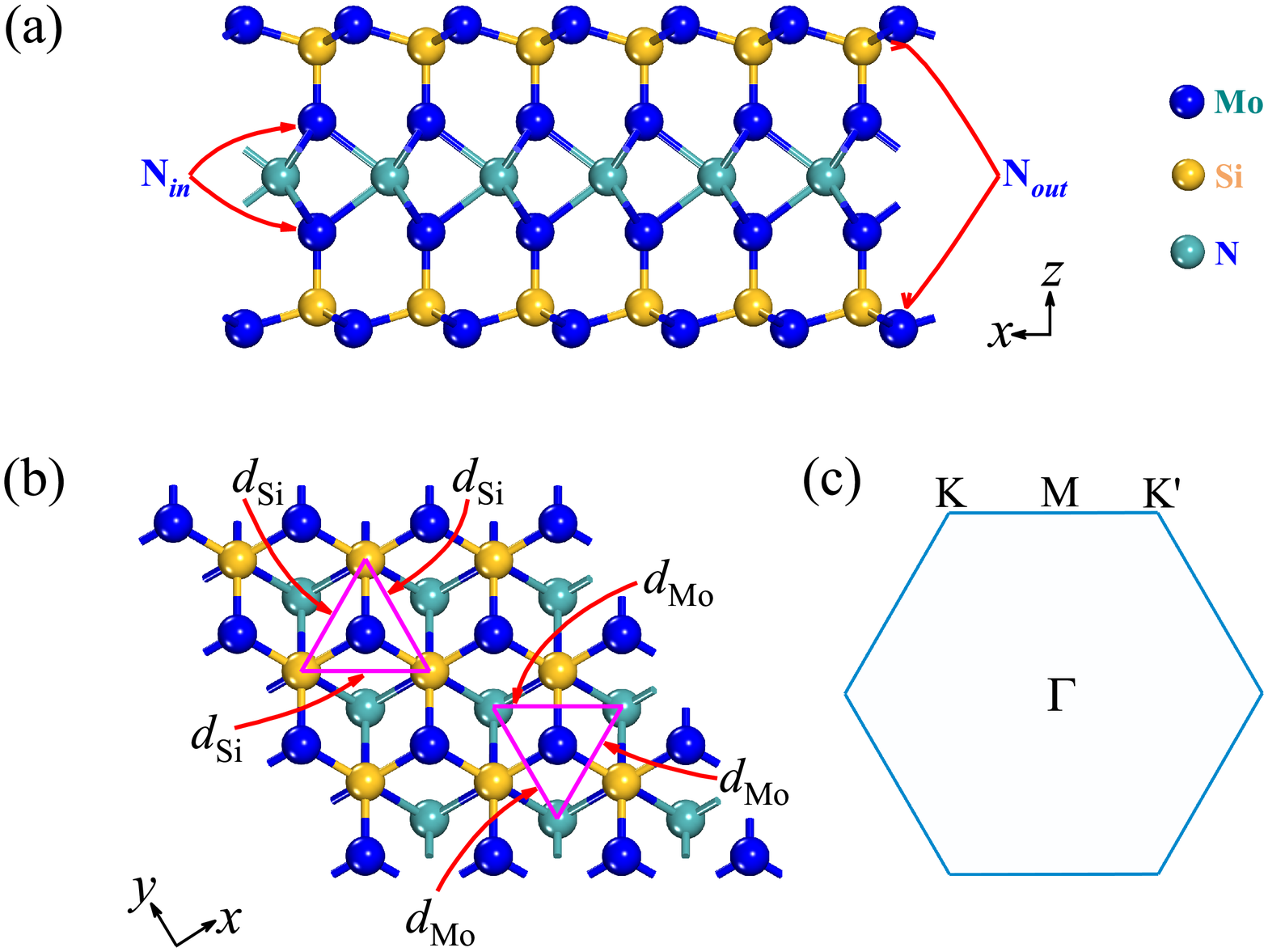}
	\caption{(Color online) Views of intrinsic MoSi$_2$N$_4$ monolayer as seen from (a) side and (b) top; (c) 2D 1st Brillouin zone. (N$_{out}$ and N$_{in}$ represent the outer and inner N atomic layers, respectively; $d$$_{Si}$ denotes the Si-Si distances nearest to the O-dopant or N-vacancy in the N$_{out}$; $d$$_{Mo}$ marks the Mo-Mo distances nearest to the O-dopant or N-vacancy in the N$_{in}$.)}
	\label{FigCellBz}
\end{figure*}

Considering that there are two types of N atomic layers according to the symmetry of septuple-atomic-layered MoSi$_2$N$_4$ monolayer, namely the outer and inner layers, which were respectively marked with N$_{out}$ and N$_{in}$ in Fig. \ref{FigCellBz}(a) for the convenience of description.
For the substitutional doping type of defects, the O atom is treated as dopant to substitute for one N atom in the N$_{out}$ and N$_{in}$ respectively, which were respectively denoted as O$_N$$_{out}$ and O$_N$$_{in}$.
The vacancy type of defects come from the removal of one N atom in the N$_{out}$ and N$_{in}$ respectively, which were respectively labeled as V$_N$$_{out}$ and V$_N$$_{in}$.
Therefore there are four types of defects.
The concentrations of these defects were all set to 6.25\%, corresponding to the 2 $\times$ 2 $\times$ 1 super-cell.
Throughout our calculations, the plane-wave pseudopotential method \cite{RevModPhys.64.1045} within the framework of DFT implemented in the CASTEP code \cite{CASTEP} was adopted.
Norm-conserving pseudopotentials were applied to describe the interactions between ion-cores and valence electrons, and the generalized gradient approximation (GGA) functional following the Perdew-Burke-Ernzerhof (PBE) \cite{PhysRevLett.77.3865} scheme was employed to paint the electronic exchange-correlation potential.
In order to minimize the interaction between the periodic layers along the c-axis direction, the vacuum slab of more than 24 {\AA} between layers was adopted.
All geometric structures were fully relaxed until the maximum force on each atom is less than 0.01 eV/{\AA}.
The cut-off energy for plan-wave expansion was set to 1100 eV and the Monkhorst-Pack \cite{PhysRevB.13.5188} \textbf{k}-point grid of 11 $\times$ 11 $\times$ 1 in the Brillouin zone was applied to both the structural relaxation and self-consistent field (SCF) process with a convergence threshold of 1.0 $\times$ 10$^{-7}$ eV/atom.

Figs. \ref{FigCellBz}(a) and \ref{FigCellBz}(b) present the side and top views of intrinsic MoSi$_2$N$_4$ monolayer respectively, which belongs to the hexagonal lattice with the P-6M2 space group (No. 187).
The lattice parameter $a$ ($a$ = $b$) of the unit cell, monolayer height $h$ and cohesive energy $E$$_{coh}$ for intrinsic MoSi$_2$N$_4$ monolayer were calculated first to examine the validity of the DFT-PBE method employed in this letter.
The calculated $a$, $h$ and $E$$_{coh}$ are respectively 2.932 {\AA}, 7.075 {\AA} and -8.573 eV/atom, which agree well with previous values \cite{Hong670,guo2020intrinsic,bafekry2020mosi2n4,li2020valleydependent,cao2020twodimensional,zhong2020straininduced}, indicating the accuracy of this method.
Tab. \ref{Tab1} summarizes the Si-Si distances (labeled as $d$$_{Si}$ in Fig. \ref{FigCellBz}(b)) nearest to the O-dopant or N-vacancy in the N$_{out}$, Mo-Mo distances (labeled as $d$$_{Mo}$ in Fig. \ref{FigCellBz}(b)) nearest to the O-dopant or N-vacancy in the N$_{in}$, lattice parameter $a'$ ($a'$ = $b'$) of the 2 $\times$ 2 $\times$ 1 super-cell and monolayer height $h$ for intrinsic and four defective systems.
It is clear from this table that the $d$$_{Si}$ in the system with O$_N$$_{out}$ and $d$$_{Mo}$ in the system with O$_N$$_{in}$ increase, whereas the $d$$_{Si}$ in the system with V$_N$$_{out}$ and $d$$_{Mo}$ in the system with V$_N$$_{in}$ decrease, compared with those of intrinsic system.
Moreover, a little bigger for the system with O$_N$$_{out}$ or O$_N$$_{in}$ toward the lattice parameter $a'$ than intrinsic system, yet the reduction of $a'$ was found in the system with V$_N$$_{out}$ or V$_N$$_{in}$.
In addition, all the four types of defects can also have an effect on the height $h$ of MoSi$_2$N$_4$ monolayer.
From Tab. \ref{Tab1}, the height $h$ of the system with O$_N$$_{out}$ or V$_N$$_{out}$ is greater than that of intrinsic system.
Nevertheless, the $h$ becomes smaller for the system with O$_N$$_{in}$ or V$_N$$_{in}$.
It is shown that the increase or decrease of $h$ may be independent of the implementation of O-doping or N-vacancy, which is only related to which N atomic layer the O-doping or N-vacancy actualized in.
Furthermore, the monolayer height will increase for O-doping or N-vacancy actualized in the N$_{out}$; on the contrary, it will decrease for that implemented in the N$_{in}$.

\begin{figure*}
	\includegraphics[height=11.0cm]{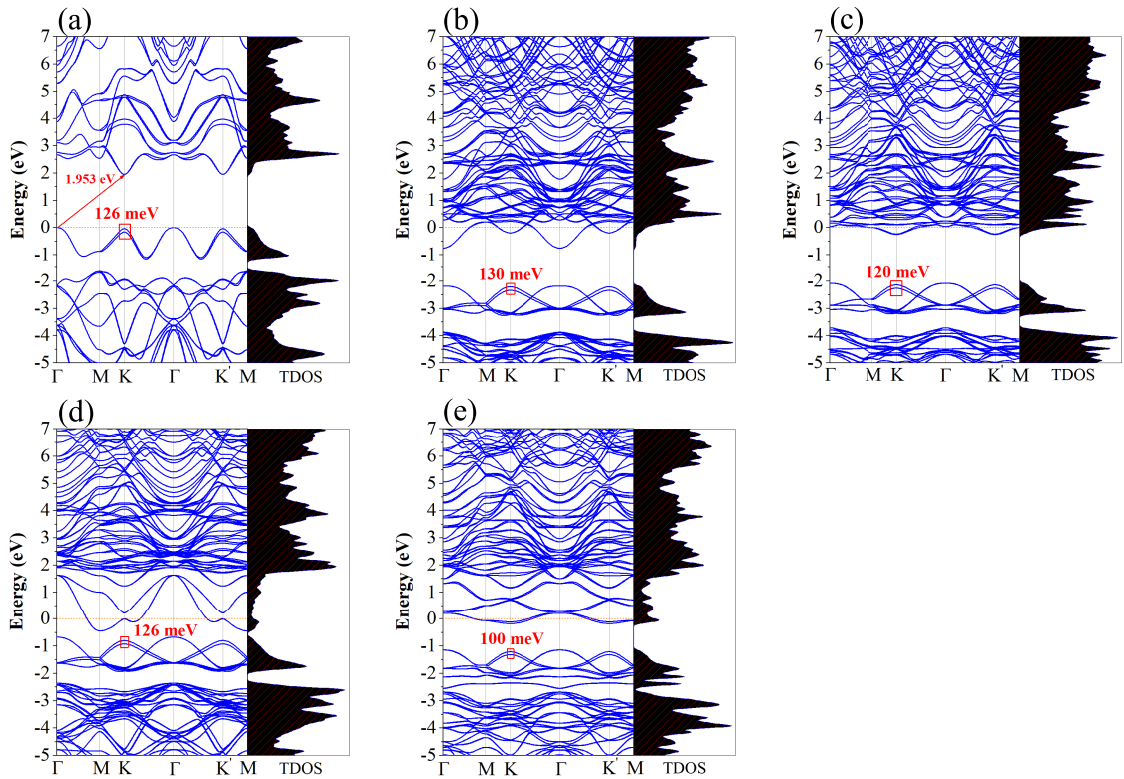}
	\caption{(Color online) Electronic band structures and corresponding TDOS incorporating SOC in a representative window around the Fermi level (is set to 0 eV) of (a) intrinsic and four systems with (b) O$_N$$_{out}$, (c) O$_N$$_{in}$, (d) V$_N$$_{out}$ and (e) V$_N$$_{in}$. (Red arrow and frames highlight the indirect band gap and spin-orbit splittings in the VB at K point, respectively.)}
	\label{FigBandTDOS}
\end{figure*}

The calculated electronic band structures along the high symmetric paths in the 1st Brillouin zone (Fig. \ref{FigCellBz}(c)) and corresponding total density of states (TDOS) incorporating SOC of intrinsic and four defective systems are displayed in Figs. \ref{FigBandTDOS}(a) $\thicksim$ \ref{FigBandTDOS}(e), respectively.
For the intrinsic system, the valence band maximum (VBM) is located at the $\Gamma$ point whereas the conduction band minimum (CBM) is located at the K point, indicating an indirect band gap semiconductor characteristic (see Fig. \ref{FigBandTDOS}(a)).
The calculated band gap value is 1.953 eV, which is in good accordance with previous result (1.940 eV) evaluated by Hong \textit{et al}. \cite{Hong670} through Tauc plot analysis.
Furthermore, the calculated spin-orbit splitting in the VB at K point is 126 meV, which is highly consistent with previous data available \cite{Hong670,guo2020intrinsic,guo2020tuning,li2020valleydependent}.
Interestingly enough, a transition from semiconductor to metal induced by each of the four types of defects was observed in MoSi$_2$N$_4$ monolayer.
In other words, for the system with O$_N$$_{out}$, O$_N$$_{in}$, V$_N$$_{out}$ or V$_N$$_{in}$, it is quite evident that each of them exhibits metallic characteristic with two dispersive energy band passing over the Fermi level (see Figs. \ref{FigBandTDOS}(b) $\thicksim$ \ref{FigBandTDOS}(e)).
Moreover, the spin-orbit splitting values are respectively 130, 120, 126 and 100 meV for the four systems with O$_N$$_{out}$, O$_N$$_{in}$, V$_N$$_{out}$ and V$_N$$_{in}$.
Compared with the splitting value of intrinsic system, there is a tiny increasement for that of the system with O$_N$$_{out}$.
But for the system with O$_N$$_{in}$ or V$_N$$_{in}$, the splitting value decreases.
It is suggested that all the O$_N$$_{out}$, O$_N$$_{in}$ and V$_N$$_{in}$ have a marginal effect on the spin-orbit coupling in MoSi$_2$N$_4$ monolayer.

We detected the Mulliken overlap populations \cite{PhysRevB.54.16317} between the center N atom in the N$_{out}$ and its three Si ligands, the center N atom in the N$_{in}$ and its three Mo ligands for the intrinsic system further.
It is found that the average overlap populations of the three N-Si and three N-Mo bonds are 1.45 and 0.94, respectively.
The absolute values of the overlap populations reflect the level of covalent properties for the N-Si and N-Mo bonds, and also mean the extent of the electron sharing between them.
However, the average overlap populations of the three O-Si bonds in the system with O$_N$$_{out}$ and three O-Mo bonds in the system with O$_N$$_{in}$ are -0.25 and -0.02, respectively.
It is shown that both the O-Si and O-Mo bonds exhibit antibonding states, implying there exist repulsive interactions, which may be the causes for the larger $d$$_{Si}$ in the system with O$_N$$_{out}$, $d$$_{Mo}$ in the system with O$_N$$_{in}$ and $a'$ in the two systems with O$_N$$_{out}$ and O$_N$$_{in}$.
Conversely, the removal of one N atom in the N$_{out}$ or N$_{in}$ from the intrinsic MoSi$_2$N$_4$ monolayer causes structural constrictions of the three Si atoms or three Mo atoms surrounding the N-vacancy.
Three bonding electrons in the system with V$_N$$_{out}$ or four bonding electrons in the system with V$_N$$_{in}$ are removed while the formation of N-vacancy.
Such a close neighbor of the vacancy of the Si atoms or Mo atoms will appear three effective charges of dangling bonds.
For the intrinsic system, the average charges of the three Si atoms directly bonding with the center N atom in the N$_{out}$ and three Mo atoms directly bonding with that in the N$_{in}$ are respectively 1.03\emph{e} and 0.06\emph{e}, according to the calculated Mulliken atomic populations.
Nonetheless, the average charge of 0.64\emph{e} was observed on the three Si atoms nearest to the N-vacancy in the system with V$_N$$_{out}$, and the average charge of the three Mo atoms nearest to the N-vacancy in the system with V$_N$$_{in}$ is -0.06\emph{e}.
It is shown that the electron density residing on the three Si atoms or three Mo atoms is increased.
Such charge redistribution can be seen as the micro-explanation for the reductions of $d$$_{Si}$ in the system with V$_N$$_{out}$, $d$$_{Mo}$ in the system with V$_N$$_{in}$ and $a'$ in the two systems with V$_N$$_{out}$ and V$_N$$_{in}$.

\begin{figure*}
	\includegraphics[height=12.0cm]{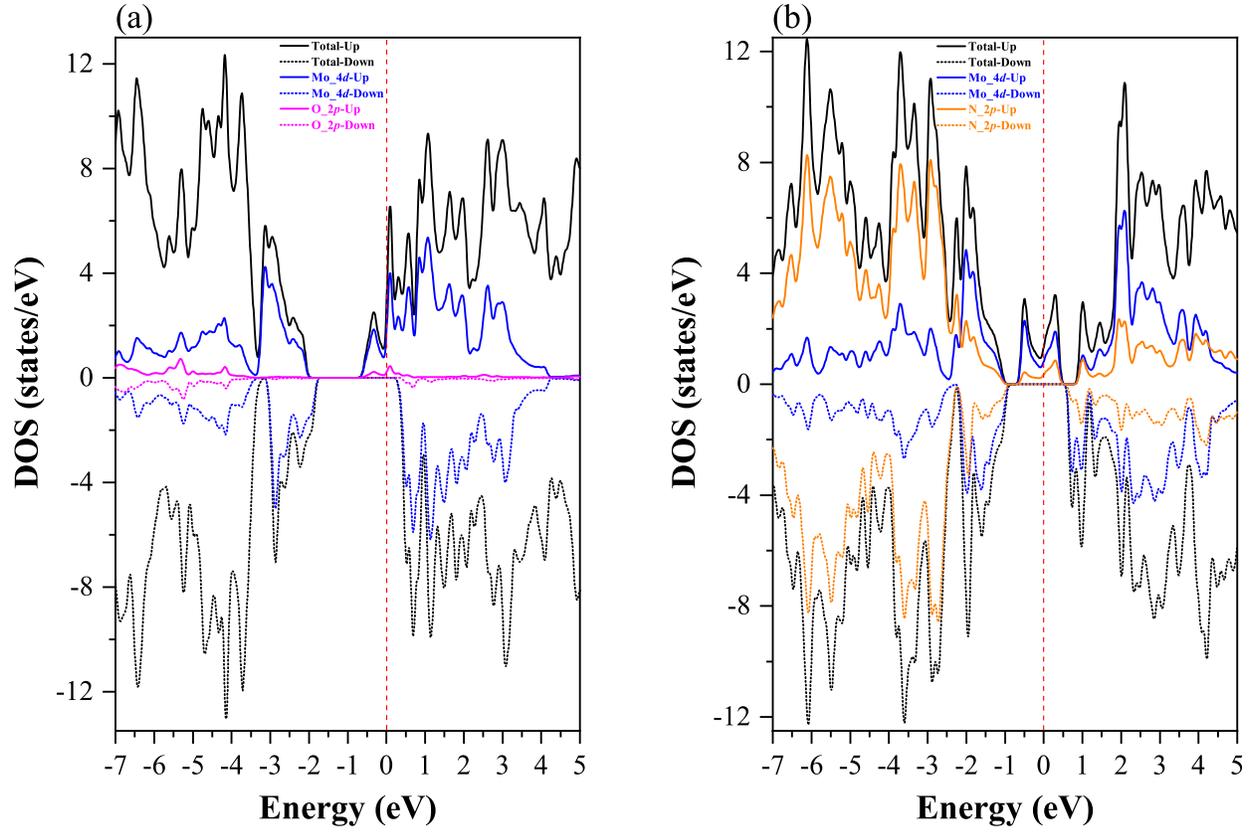}
	\caption{(Color online) DOS projected on (a) Mo and O atoms in the system with O$_N$$_{in}$ and (b) Mo and N atoms in the system with V$_N$$_{in}$.}
	\label{FigO2V2OrbPDOS}
\end{figure*}

The spin polarized calculations without SOC and given initial magnetic configuration were performed to analyze the magnetic behavior.
According to the calculated Mulliken atomic populations, the total magnetic moments $m$$_{tot}$ of intrinsic and four defective systems are also tabulated in Tab. \ref{Tab1}.
From this table, the intrinsic MoSi$_2$N$_4$ monolayer possesses no magnetic moment, which is consistent with previous result calculated by Hong \textit{et al}. \cite{Hong670}, showing that this calculation method is reliable.
For the four defective systems, the total magnetic moment of the system with O$_N$$_{out}$ or V$_N$$_{out}$ is also zero.
Nevertheless, a $m$$_{tot}$ of 1.000$\mu$$_B$ was observed in the system with O$_N$$_{in}$ or V$_N$$_{in}$.
As displayed in Fig. \ref{FigO2V2OrbPDOS}(a), the partial density of states (PDOS) of Mo and O atoms implies that the total magnetic moment mainly originates from the strong coupling between the 4\emph{d} states of Mo and 2\emph{p} state of O for the system with O$_N$$_{in}$.
For the system with V$_N$$_{in}$, the PDOS of Mo and N atoms suggests that the total magnetic moment primarily comes from the strong coupling between the 4\emph{d} states of Mo and 2\emph{p} state of N, as shown in Fig. \ref{FigO2V2OrbPDOS}(b).

\begin{figure*}
	\includegraphics[height=9.0cm]{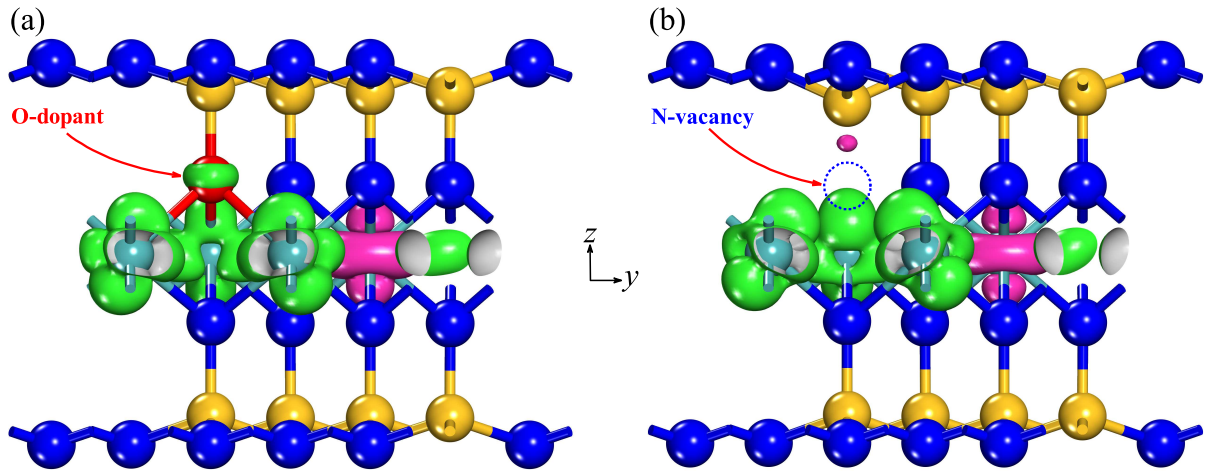}
	\caption{(Color online) Side views of net spin density at an isovalue of 5 $\times$ 10$^{-2}$ \emph{e}/{\AA}$^{3}$ for the two systems with (a) O$_N$$_{in}$ and (b) V$_N$$_{in}$. (The up-spin and down-spin states are denoted as green and magenta, respectively.)}
	\label{FigO2V2SpinD}
\end{figure*}

\begin{figure*}
	\includegraphics[height=9.0cm]{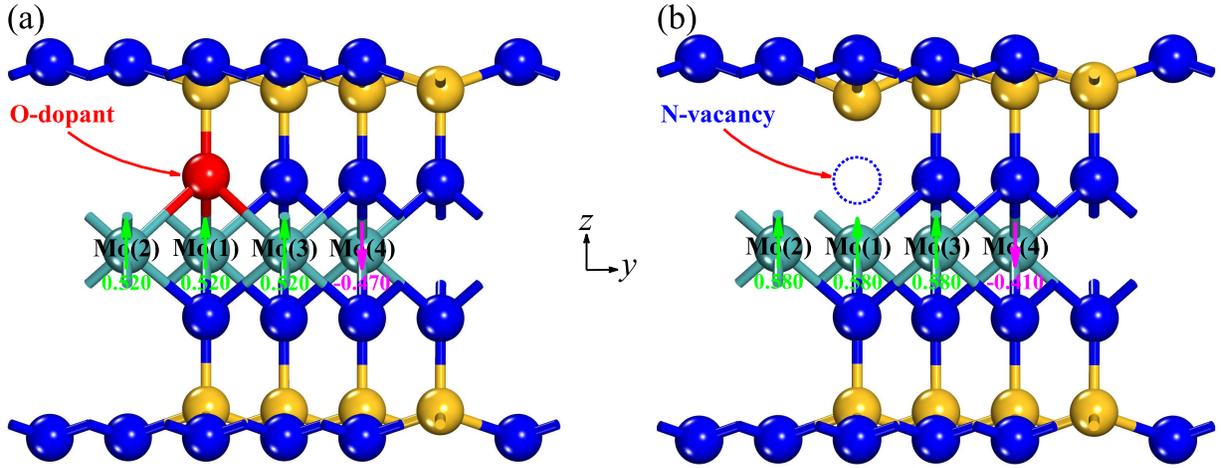}
	\caption{(Color online) The detailed magnetic moment arrangements within a super-cell in the two systems with (a) O$_N$$_{in}$ and (b) V$_N$$_{in}$.}
	\label{FigO2V2Spin}
\end{figure*}

We inspected the local magnetic moments of the two systems with O$_N$$_{in}$ and V$_N$$_{in}$ further.
Figs. \ref{FigO2V2SpinD}(a) and \ref{FigO2V2SpinD}(b) show the side views of net spin density for the two systems with O$_N$$_{in}$ and V$_N$$_{in}$, respectively.
As shown in these figures, the total magnetic moment is notably contributed from the local magnetic moments residing on the Mo atoms.
According to these local magnetic moments, the magnetic structures for both of these systems can be demonstrated.
Figs. \ref{FigO2V2Spin}(a) and \ref{FigO2V2Spin}(b) present the detailed magnetic moment arrangements within a super-cell in the two systems with O$_N$$_{in}$ and V$_N$$_{in}$, respectively.
From these figures, the local magnetic moments of Mo(1), Mo(2) and Mo(3) atoms around the O-dopant or N-vacancy are equivalent to each other, exhibiting up-spin states, and the magnitudes are 0.520$\mu$$_B$ and 0.580$\mu$$_B$ per Mo atom for the two systems with O$_N$$_{in}$ and V$_N$$_{in}$, respectively.
However, the remaining one Mo(4) atom possesses a local magnetic moment with down-spin state, and the magnitudes are -0.470$\mu$$_B$ and -0.410$\mu$$_B$ per Mo atom for the two systems with O$_N$$_{in}$ and V$_N$$_{in}$, respectively.
It can be concluded that such magnetic moment arrangement displays a magnetic structure with ferrimagnetic (FRM) ordering for the system with O$_N$$_{in}$ or V$_N$$_{in}$.

\begin{figure*}
	\includegraphics[height=12.0cm]{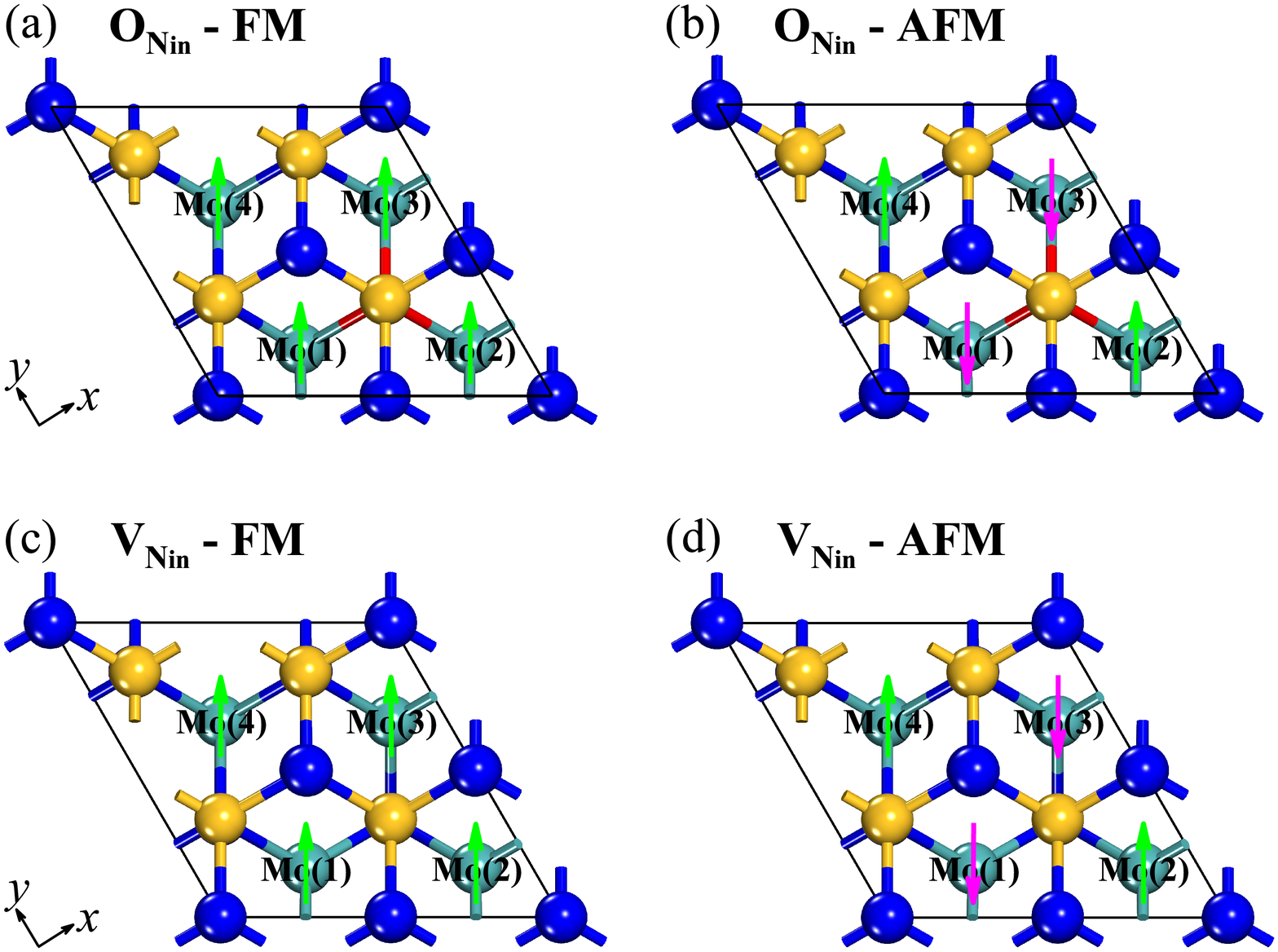}
	\caption{(Color online) The given initial magnetic configurations of the two systems with O$_N$$_{in}$ and V$_N$$_{in}$: two ferromagnetic states (a, c) and two antiferromagnetic states (b, d).}
	\label{FigFMAFM}
\end{figure*}

In order to verify the magnetic structure further, two different initial magnetic configurations within the 2 $\times$ 2 $\times$ 1 super-cells of the system with O$_N$$_{in}$ or V$_N$$_{in}$ were given, namely, one ferromagnetic (FM) and one antiferromagnetic (AFM) initial states (see Figs. \ref{FigFMAFM}(a) $\thicksim$ \ref{FigFMAFM}(d)).
We found that whatever the initial magnetic configuration is, the system with O$_N$$_{in}$ or V$_N$$_{in}$ becomes a ferrimagnetic state finally which is the same as that (see Figs. \ref{FigO2V2Spin}(a) and \ref{FigO2V2Spin}(b)) obtained using spin polarized calculations without given initial magnetic configuration.
It is indicated that the system with O$_N$$_{in}$ or V$_N$$_{in}$ should possess a ferrimagnetic ground state.

Next, the effect of biaxial strain on the magnetic properties of the two systems with O$_N$$_{in}$ and V$_N$$_{in}$ was discussed.
Here, the biaxial strain $\varepsilon$ is defined as
\begin{equation}
\varepsilon = \frac{a'-a'_0}{a'_0} \times 100\%,
\label{strain}
\end{equation}
where $a'$ and $a'$$_0$ are the strained and unstrained lattice parameters, respectively.
The positive (negative) $\varepsilon$ represents the tensile (compressive) strain, and a strain range from -10\% to 10\% was applied to both of these systems.
Tab. \ref{TabOin} and Tab. \ref{TabVin} list the calculated total magnetic moment $m$$_{tot}$, local magnetic moments residing on the Mo atoms $m$$_{Mo}$$_{(1)}$, $m$$_{Mo}$$_{(2)}$, $m$$_{Mo}$$_{(3)}$ and $m$$_{Mo}$$_{(4)}$ of the two systems with O$_N$$_{in}$ and V$_N$$_{in}$ under different strains, respectively.
From Tab. \ref{TabOin}, it is obvious that the system with O$_N$$_{in}$ exhibits a ferrimagnetic state within the range of -3\% $\thicksim$ 10\%.
The variation of $m$$_{tot}$ is negligible as strain increases from -3\% to 10\%, indicating that the $m$$_{tot}$ is less sensitive to biaxial strain.
Nevertheless, the $m$$_{Mo}$$_{(1)}$, $m$$_{Mo}$$_{(2)}$, $m$$_{Mo}$$_{(3)}$ and $m$$_{Mo}$$_{(4)}$ are all increased.
It is worth noting that the local magnetic moments residing on the Mo(1), Mo(2) and Mo(3) atoms are still equivalent to each other, which may be associated with the unbroken lattice symmetry under biaxial strain.
Moreover, the $m$$_{tot}$, $m$$_{Mo}$$_{(1)}$, $m$$_{Mo}$$_{(2)}$, $m$$_{Mo}$$_{(3)}$ and $m$$_{Mo}$$_{(4)}$ are all zero within the range of -10\% $\thicksim$ -5\%, indicating this system shows a paramagnetic state.
We found that the $m$$_{tot}$, $m$$_{Mo}$$_{(1)}$, $m$$_{Mo}$$_{(2)}$, $m$$_{Mo}$$_{(3)}$ and $m$$_{Mo}$$_{(4)}$ at -4\% strain are all difficult to calculate, cannot be obtained, and thus the data at such strain are not given in Tab. \ref{TabOin}.
Therefore we suggest that there exists a magnetic phase transition between ferrimagnetic and paramagnetic states around -4\% strain.
For the system with V$_N$$_{in}$, it is clear from Tab. \ref{TabVin} that it presents a ferrimagnetic state within the range of -9\% $\thicksim$ 10\%.
Analogously, the $m$$_{tot}$ is also less sensitive to biaxial strain, the $m$$_{Mo}$$_{(1)}$, $m$$_{Mo}$$_{(2)}$, $m$$_{Mo}$$_{(3)}$ and $m$$_{Mo}$$_{(4)}$ are all increased, the local magnetic moments of Mo(1), Mo(2) and Mo(3) atoms are still equivalent to each other, as strain increases from -9\% to 10\%.
However, the $m$$_{tot}$, $m$$_{Mo}$$_{(1)}$, $m$$_{Mo}$$_{(2)}$, $m$$_{Mo}$$_{(3)}$ and $m$$_{Mo}$$_{(4)}$ are all zero at -10\% strain, suggesting this system displays a paramagnetic state.
We calculated these magnetic moments at -11\% strain further and found that all of them are also zero.
It is shown that there also exists a magnetic phase transition between ferrimagnetic and paramagnetic states within the range of -10\% $\thicksim$ -9\%.

In conclusion, we systematically investigated the structural, electronic and magnetic properties of MoSi$_2$N$_4$ monolayer induced by the four types of defect (O$_N$$_{out}$, O$_N$$_{in}$, V$_N$$_{out}$ and V$_N$$_{in}$), using first-principles density functional theory calculations.
Our results reveal that all the four types of defects lead to structural distortions around the O-dopant or N-vacancy, and thereby change the lattice parameter $a'$ and monolayer height $h$.
Specifically, O$_N$$_{out}$ or O$_N$$_{in}$ increases the lattice parameter, but V$_N$$_{out}$ or V$_N$$_{in}$ is on the contrary.
O$_N$$_{out}$ or V$_N$$_{out}$ increases the monolayer height, whereas the height decreases for O$_N$$_{in}$ or V$_N$$_{in}$.
Each of the four types of defects has a fundamental effect on the electronic properties of MoSi$_2$N$_4$ monolayer, which can induce a transition from semiconductor to metal.
O$_N$$_{in}$ or V$_N$$_{in}$ plays a vital role in the occurrence of a transition from non-magnetism to ferrimagnetism in MoSi$_2$N$_4$ monolayer.
Subsequently, the effect of biaxial strain on the magnetic properties of the two systems with O$_N$$_{in}$ and V$_N$$_{in}$ was investigated.
It is found that the total magnetic moments are less sensitive to biaxial strain, the local magnetic moments residing on the Mo atoms are increased for the two systems with O$_N$$_{in}$ and V$_N$$_{in}$, as strain increases from -3\% to 10\% and from -9\% to 10\%, respectively.
Furthermore, the magnetic phase transitions between ferrimagnetic and paramagnetic states were found to occur around -4\% strain and within the range of -10\% $\thicksim$ -9\% for the two systems with O$_N$$_{in}$ and V$_N$$_{in}$, respectively.
This study may provide a guidance for the application of MoSi$_2$N$_4$ monolayer in the spintronic and magnetic materials.



\nocite{*}
\bibliography{aipsamp}


\clearpage

\begin{table*}
\caption{The Si-Si distances ($d$$_{Si}$) nearest to the O-dopant or N-vacancy in the N$_{out}$, Mo-Mo distances ($d$$_{Mo}$) nearest to the O-dopant or N-vacancy in the N$_{in}$, lattice parameters $a'$ ($a'$ = $b'$) of the 2 $\times$ 2 $\times$ 1 super-cell, monolayer height $h$ and total magnetic moment $m$$_{tot}$ for intrinsic and four defective systems.}
\begin{ruledtabular}
\begin{tabular}{cccccc}
Systems        & $d$$_{Si}$     & $d$$_{Mo}$            & $a'$         & $h$     &    $m$$_{tot}$                \\
               &  ({\AA})       &  ({\AA})              & ({\AA})      & ({\AA}) &  ($\mu$$_B$/super-cell)       \\  \hline
               &                &                       &              &         &                                \\
Intrinsic      &   2.932        &  2.932                & 5.864        & 7.075   &        0.000                    \\
               &                &                       &              &         &                                 \\
O$_N$$_{out}$  &   3.060        &  ---                  & 5.887        & 7.092   &        0.000                     \\
               &                &                       &              &         &                                  \\
O$_N$$_{in}$   &    ---         &  2.989                & 5.884        & 7.067   &        1.000                     \\
               &                &                       &              &         &                                    \\
V$_N$$_{out}$  &   2.856        &  ---                  & 5.856        & 7.103   &        0.000                      \\
               &                &                       &              &         &                                    \\
V$_N$$_{in}$   &    ---         &  2.917                & 5.850        & 7.065   &        1.000			              \\
               &                &                       &              &         &
\end{tabular}
\end{ruledtabular}
\label{Tab1}
\end{table*}

\begin{table*}
\caption{The calculated total magnetic moment $m$$_{tot}$, local magnetic moments residing on the Mo atoms $m$$_{Mo}$$_{(1)}$, $m$$_{Mo}$$_{(2)}$, $m$$_{Mo}$$_{(3)}$ and $m$$_{Mo}$$_{(4)}$ of the system with O$_N$$_{in}$ under different strains.}
\begin{ruledtabular}
\begin{tabular}{cccccc}
Strain & $m$$_{tot}$                        & $m$$_{Mo}$$_{(1)}$              & $m$$_{Mo}$$_{(2)}$             & $m$$_{Mo}$$_{(3)}$           & $m$$_{Mo}$$_{(4)}$     \\
	   & ($\mu$$_B$/super-cell)             & ($\mu$$_B$/atom)                & ($\mu$$_B$/atom)               & ($\mu$$_B$/atom)             & ($\mu$$_B$/atom)       \\   \hline
	   &                                    &                                 &                                &                              &                         \\	
-10\%  & 0.000                              & 0.000                           & 0.000                          & 0.000                        & 0.000                   \\
       &                                    &                                 &                                &                              &                         \\
-8\%   & 0.000                              & 0.000                           & 0.000                          & 0.000                        & 0.000                  \\
       &                                    &                                 &                                &                              &                         \\
-6\%   & 0.000                              & 0.000                           & 0.000                          & 0.000                        & 0.000                    \\
       &                                    &                                 &                                &                              &                         \\
-5\%   & 0.000                              & 0.000                           & 0.000                          & 0.000                        & 0.000                   \\
       &                                    &                                 &                                &                              &                          \\
-4\%   & ---                                & ---                             & ---                            & ---                          & ---                     \\
       &                                    &                                 &                                &                              &                          \\
-3\%   & 0.997                              & 0.330                           & 0.330                          & 0.330                        & -0.120                   \\
       &                                    &                                 &                                &                              &                          \\
-2\%   & 1.000                              & 0.340                           & 0.340                          & 0.340                        & -0.140                    \\
       &                                    &                                 &                                &                              &                          \\
0      & 1.000                              & 0.520                           & 0.520                          & 0.520                        & -0.470                    \\
       &                                    &                                 &                                &                              &                           \\
2\%    & 0.999                              & 0.560                           & 0.560                          & 0.560                        & -0.510                    \\
       &                                    &                                 &                                &                              &                           \\
4\%    & 0.969                              & 0.590                           & 0.590                          & 0.590                        & -0.590                    \\
       &                                    &                                 &                                &                              &                           \\
6\%    & 0.975                              & 0.680                           & 0.680                          & 0.680                        & -0.820                     \\
       &                                    &                                 &                                &                              &                           \\
8\%    & 1.000                              & 0.880                           & 0.880                          & 0.880                        & -1.340                    \\
       &                                    &                                 &                                &                              &                            \\
10\%   & 1.000                              & 1.100                           & 1.100                          & 1.100                        & -1.920                     \\
       &                                    &                                 &                                &                              &
\end{tabular}
\end{ruledtabular}
\label{TabOin}
\end{table*}

\begin{table*}
\caption{The calculated total magnetic moment $m$$_{tot}$, local magnetic moments residing on the Mo atoms $m$$_{Mo}$$_{(1)}$, $m$$_{Mo}$$_{(2)}$, $m$$_{Mo}$$_{(3)}$ and $m$$_{Mo}$$_{(4)}$ of the system with V$_N$$_{in}$ under different strains.}
\begin{ruledtabular}
\begin{tabular}{cccccc}
Strain & $m$$_{tot}$                        & $m$$_{Mo}$$_{(1)}$              & $m$$_{Mo}$$_{(2)}$             & $m$$_{Mo}$$_{(3)}$           & $m$$_{Mo}$$_{(4)}$     \\
	   & ($\mu$$_B$/super-cell)             & ($\mu$$_B$/atom)                & ($\mu$$_B$/atom)               & ($\mu$$_B$/atom)             & ($\mu$$_B$/atom)       \\   \hline
	   &                                    &                                 &                                &                              &                         \\
-10\%  & 0.000                              & 0.000                           & 0.000                          & 0.000                        & 0.000                \\
	   &                                    &                                 &                                &                              &                     \\
-9\%   & 0.946                              & 0.380                           & 0.380                          & 0.380                        & -0.170              \\
       &                                    &                                 &                                &                              &                    \\
-8\%   & 0.996                              & 0.420                           & 0.420                          & 0.420                        & -0.190             \\
	   &                                    &                                 &                                &                              &                    \\
-6\%   & 1.000                              & 0.460                           & 0.460                          & 0.460                        & -0.230             \\
	   &                                    &                                 &                                &                              &                    \\
-4\%   & 1.000                              & 0.500                           & 0.500                          & 0.500                        & -0.280              \\
	   &                                    &                                 &                                &                              &                    \\
-2\%   & 1.000                              & 0.540                           & 0.540                          & 0.540                        & -0.340            \\
	   &                                    &                                 &                                &                              &                      \\
0      & 1.000                              & 0.580                           & 0.580                          & 0.580                        & -0.410             \\
	   &                                    &                                 &                                &                              &                     \\
2\%    & 1.000                              & 0.630                           & 0.630                          & 0.630                        & -0.520                \\
	   &                                    &                                 &                                &                              &                       \\
4\%    & 1.000                              & 0.710                           & 0.710                          & 0.710                        & -0.690                \\
	   &                                    &                                 &                                &                              &                     \\
6\%    & 1.001                              & 0.820                           & 0.820                          & 0.820                        & -0.970              \\
	   &                                    &                                 &                                &                              &                      \\
8\%    & 1.001                              & 1.000                           & 1.000                          & 1.000                        & -1.430               \\
	   &                                    &                                 &                                &                              &                       \\
10\%   & 1.055                              & 1.220                           & 1.220                          & 1.220                        & -1.930               \\
	   &                                    &                                 &                                &                              &
\end{tabular}
\end{ruledtabular}
\label{TabVin}
\end{table*}

\end{document}